\documentclass[superscriptaddress,floats,a4paper,floatfix,prb]{revtex4}
\usepackage{graphicx}
\usepackage{dcolumn}
\usepackage{bm}

\begin{document}

\title{Evidence for differentiation in the iron-helicoidal-chain in GdFe$_{3}$(BO$_{3}$)$_{4}$ }

\author{S.A. Klimin}
\affiliation{Material Science Center, University of
        Groningen, 9747 AG Groningen, The Netherlands.}
\affiliation{Institute of Spectroscopy, RAS, 142190, Troitsk, Moscow
        Region, Russia.}
\author{D. Fausti}
\author{A. Meetsma}
\affiliation{Material Science Center, University of
        Groningen, 9747 AG Groningen, The Netherlands.}
\author{L.N. Bezmaternykh}
\affiliation{L.V.Kirensky Institute of Physics, Siberian branch
               of RAS, Krasnoyarsk 660036, Russia.}
\author{P.H.M. van Loosdrecht}
\affiliation{Material Science Center, University of
        Groningen, 9747 AG Groningen, The Netherlands.}
\author{T.T.M. Palstra}
\affiliation{Material Science Center, University of
        Groningen, 9747 AG Groningen, The Netherlands.}


\begin{abstract}
We report on a single-crystal X-ray structure study of
GdFe$_3$(BO$_3$)$_4$ at room temperature and at T=90~K. At room
temperature GdFe$_3$(BO$_3$)$_4$ crystallizes in a trigonal space
group $R$32 (No. 155), the same as found for other members of
iron-borate family RFe$_3$(BO$_3$)$_4$. At 90 K the structure of
GdFe$_3$(BO$_3$)$_4$ transforms to the space group $P$3$_{1}$21
(No. 152). The low-temperature structure determination gives new
insight into the weakly first-order structural phase transition at
156~K and into the related Raman phonon anomalies. The presence of
two inequivalent iron chains in the low temperature structure
provides a new perspective on the interpretation of the
low-temperature magnetic properties.
\end{abstract}
\maketitle

\section{Introduction}
The family of borates RM$_3$(BO$_3$)$_4$ with a rare earth (RE) or
yttrium as R and Al, Ga, Fe, or Sc as M crystallize in the
huntite, CaMg$_3$(CO$_3$)$_4$, structure type with space group
$R$32 \cite{Juob,Belo,Camp}. The interest in this family of
crystals arises both from a fundamental point of view and from
already realized and proposed applications. Crystals of
YAl$_3$(BO$_3$)$_4$ and GdAl$_3$(BO$_3$)$_4$ doped with Nd were
widely studied during recent years and have been used in optical
devices, such as self-frequency doubling and self-frequency
summing lasers (see, e.g., Ref. \cite{Bren} and references
therein). Concentrated NdAl$_3$(BO$_3$)$_4$ crystals are efficient
media for minilasers \cite{che}.
\\ Apart from the interesting optical properties arising chiefly 
from the  lack of inversion
symmetry, the 'sub-family' of RFe$_3$(BO$_3$)$_4$ also attracts
considerable attention due to their structure peculiarities. The
presence of magnetic order at temperatures less then 37 K was
attributed to magnetic Fe-Fe or Fe-O-Fe interactions inside quasi
one-dimensional (1D) iron chains \cite{Camp,bala}. Recent works
\cite{levi,bala}, focusing on low temperature magnetism in
GdFe$_3$(BO$_3$)$_4$, revealed two magnetic phase transitions. The
second-order magnetic ordering phase transition at
$T_{\mathrm{N1}}=$37~K (antiferromagnetic ordering of Fe atoms) is
followed by a first-order spin-reorientational phase transition at
$T_{\mathrm{N2}}=$10~K. Additionally specific heat and Raman
measurements on single crystals \cite{levi} of
GdFe$_3$(BO$_3$)$_4$ revealed a weakly first-order structural
phase transition at $T_{\mathrm{s}}=$156~K. This structural phase
transition is observed in almost all members of the RE
ferro-borates family \cite{hina,levi,loos}. Recently, Hinatsu et
al. \cite{hina} have shown that DyFe$_3$(BO$_3$)$_4$ undergoes
such structural phase transition at 340~K, by measuring the
temperature dependence of the lattice parameters on powder
samples. Here too, a peak in the specific heat was observed at
this temperature. Similar peaks in specific heat vs temperature
dependence were found for other heavy RE (R= Eu-Ho, Gd, Tb)
ferroborates. These peaks were ascribed to structural phase
transitions. The transition temperatures $T_{\mathrm{s}}$ were
found to depend linearly on the ionic radius of the RE. To date,
there are no single crystal data available nor has the
low-temperature (LT) space group been determined.

The high temperature (HT) $R$32 structure of powder
RFe$_3$(BO$_3$)$_4$ compounds was first determined in Ref.
\cite{Juob} for R = La, Nd, Sm-Ho, and Y. X-ray experiments on
single crystals with R=Nd$_{0.5}$Bi$_{0.5}$ \cite{Belo} and R=La,
Nd, and Y$_{0.5}$Bi$_{0.5}$ \cite{Camp} confirmed this structure.
Moreover, Raman measurements for LaFe$_3$(BO$_3$)$_4$ \cite{alic},
NdFe$_3$(BO$_3$)$_4$ \cite{alic,loos} and GdFe$_3$(BO$_3$)$_4$
 \cite{loos} are in good agreement with group theoretical analysis,
based on the $R$32 structure.

Summarizing, whereas the HT crystal structure of most of the
ferroborates is known, the LT space group, lattice parameters, and
atomic positions are unknown. In this work we report on a
single-crystal X-ray diffraction structure study of
gadolinium-iron borate at room temperature (RT) and at 90~K . It
is found that the LT-structure has the $P$3$_1$21 symmetry. Two
nonequivalent spin chains exist in the LT-structure which gives
new insight in the low-temperature magnetic properties. The
$P$3$_1$21 symmetry of the LT-phase also confirms the
interpretation of the reported IR absorption by crystal-field
transitions \cite{levi,chuk} and leads to a better understanding
of some peculiarities observed in Raman spectra \cite{levi,loos}.

\section{Experiment and result of structure determination}

Crystals of GdFe$_3$(BO$_3$)$_4$ were grown using a
K$_{2}$Mo$_{3}$O$_{10}$ - based flux, as described in Ref.
\cite{bala}. Big transparent single crystals of gadolinium
ferroborates were light green in color and had a good optical
quality. A block-shaped crystal ('broken-fragment') with the
dimensions of 0.22 $\times$ 0.15 $\times$ 0.11~mm$^{3}$~was
mounted on top of a glass fiber and aligned on a Bruker
\cite{Bru1} SMART APEX CCD diffractometer. The crystal was cooled
to 90(1)~K using a Bruker KRYOFLEX. Intensity measurements were
performed using graphite monochromated Mo-K$\bar{\alpha}$
radiation. Generator settings were 50~KV/ 40~mA. SMART \cite{Bru1}
was
 used for preliminary determination of the unit cell constants and data
collection control. The intensities of reflections of a hemisphere
were collected by a combination of 6 sets of exposures (frames).
Each set had a different angle for the crystal and each exposure
covered a range of $0.3^{\circ}$  in $\omega$. A total of 3600
frames were collected with an exposure time of 10.0 seconds per
frame. The overall data collection time was 16 hours. Data
integration and global cell refinement was performed with the
program SAINT. The final unit cell was obtained from the xyz
centroids of 4767 and 5439 reflections after integration for RT
and 90 K, respectively. Intensity data were corrected for Lorentz
and polarization effects, scale variation, for decay and
absorption (an analytical absorption correction was applied), and
reduced to $F_{o}^{2}$. The program suite SHELXTL was used for
space group determination (XPREP).
\begin{figure}[htb]
\includegraphics[width=85mm]{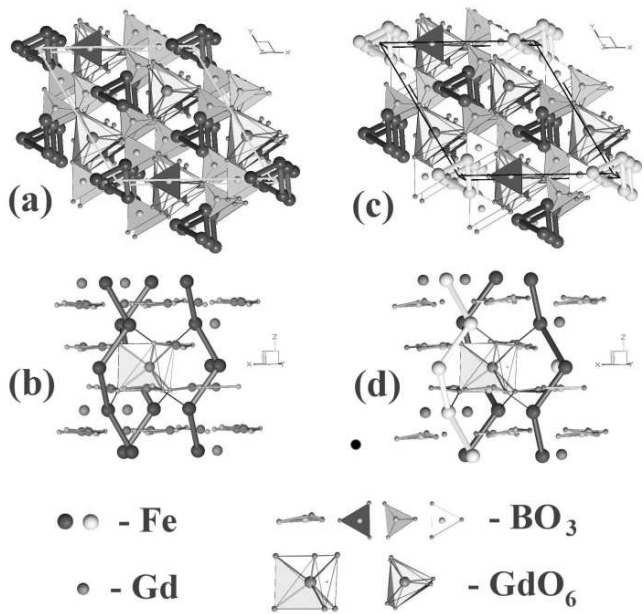}
\caption{The structure of GdFe$_{3}$(BO$_{3}$)$_{4}$ in two
different projection. The left panels (a) and (b) show the RT
structure, and the right panels (c) and (d) show the structure at
90~K. The Fe atoms are arranged in chiral chains parallel to the
\textbf{\textit{c}}-axis. Different chains are separated by
GdO$_{6}$ prisms and BO$_{3}$ groups. The unit cell outline for
$R$32 is shifted by (1/3, 1/3, 0) for comparison with the LT
$P$3$_{1}$21 structure.} \label{structure}
\end{figure}

At both temperatures of RT and 90~K the unit cell was identified
as trigonal: reduced cell calculations did not indicate any higher
metric lattice symmetry \cite{X3}. Space groups $R$32  and
$P$3$_1$21 were derived for RT and for 90~K, respectively, from
the systematic extinctions and discriminated from other candidate
space groups, which comply with the same extinctions conditions,
during the structure determination process. Examination of the
final atomic coordinates of the structure did not yield extra
crystallographic or metric symmetry elements \cite{X4,X5}.

The polarity of the structure of the crystal actually chosen was
determined by Flack's $\mathbf{x}$-refinement
\cite{flack1,flack2,flack3,Herbst1998}; refinement resulted in a
$\mathbf{x}$ value of 0.50(1), so ultimately an inversion twin is
used in the refinement.

\begin{table}[p]
\caption{Crystallographic parameters}
\tiny
\label{roomlowTcomparison}
\begin{tabular}{lll}
&&\\
\multicolumn{3}{l}{\footnotesize \textbf{a-Crystal data and details of the structure determination.}}\\
                                & Room Temperature = 293(1)~K& T = 90(1)~K \\
Moiety Formula                  &\multicolumn{2}{c}{GdFe$_{3}$(BO$_{3}$)$_{4}$}     \\
Formula Weight $(g.mol^{-1})$    &\multicolumn{2}{c}{560.04}                    \\
Crystal System                  & \multicolumn{2}{c}{Trigonal}                  \\
Space group.                    & $R$32, 155                  & $P$3$_{1}$21, 152\\
$a$\ (\AA)                         & 9.5203(6)                    & 9.5305(3)\\
$c$\ (\AA)                         & 7.5439(5)                    & 7.5479(2)\\
$V$\ (\AA$^{3}$)                   & 592.15(7)                    & 593.73(3)\\
$\Theta$ range unit cell:min.-max., &3.66-38.70;4767  & 3.66-38.55; 5439\\
 deg; reflections               &                           &       \\
Formula\_Z                       & 3                         & 3\\
Space group\_Z                   & 18                        & 6\\
Z'(=Formula\_Z/SpaceGroup\_Z)     & 1/6                       & 0.5\\
$\rho (g/cm^{3})$               & 4.712                     & 4.699\\
F(000) electrons                & 774                       & 774\\
$\mu(Mo K\alpha) (cm^{-1})$     & 137.74                    & 137.37\\

&&\\
\multicolumn{3}{l}{\footnotesize \textbf{b-Data collection.}}\\
$\lambda (MoK\bar{\alpha})$, \AA          & \multicolumn{2}{c}{0.71073}\\
Monochromator                            & \multicolumn{2}{c}{Graphite}\\
Measurements device type                &\multicolumn{2}{c}{Bruker SMART APEX,}\\
                                        &\multicolumn{2}{c}{CCD area-detector diffractometer}\\

Detector area resolution (pixels/mm)    &  \multicolumn{2}{c}{$4096\times 4096/62\times 62$~(binned 512)} \\
Measurement method                      &  $\varphi$- and $\omega$-scans     &  $\varphi$- and $\omega$-scans\\
$\theta$ range; min. max.,deg             &  3.66,  38.70     & 2.47,  38.58\\
Index ranges                            & $h:-16\rightarrow16;k:-16\rightarrow16$; & $h:-16\rightarrow16;k:-16\rightarrow16$;\\
                                        &  $l: -13\rightarrow13$    & $l: -13\rightarrow13$ \\
Min.- Max. absorption transmission factor &  0.0563-0.2519& 0.0620-0.2513\\
X-ray exposure time, h                  & 16.0 & 16.0\\
Total data                              & 4700 & 13868\\
Unique data                             & 755  & 2217\\
Data with criterion: ($F_{o}\geq 4.0\sigma(F_{o})$)    & 755  & 1956\\
$R_{int} = \sum[|F_{o}^{2} -F_{o}^{2} (mean)|]/\sum[F_{o}^{2}]$ & 0.0339 & 0.0338\\
$R_{sig} = \sum\sigma(F_{o}^{2}) / [F_{o}^{2}]$  & 0.0203 & 0.0189\\
&&\\
\multicolumn{3}{l}{\footnotesize \textbf{c-Refinement.}}\\
Number of reflections                               & 755          &2217\\
Number of refined parameters                        & 35           &95\\
Isotropic secondary-extinction coefficient, g       & 0.0371(12)   &0.0359(10)\\
Final agreement factors:   & 0.0374  &0.0452\\
$wR(F^{2}) = [\sum [w(F_{o}^{2}-F_{c}^{2})^{2}] /[w(F_{o}^{2})^{2}]]^{1/2}$ & & \\
Weighting scheme: a, b                              & 0.0272, 0.0    &0.0312, 0.0\\
$w=1/[\sigma^{2}(F_{o}^{2})+(aP)^{2}+bP]$            &               &\\
And $P=[max(F_{o}^{2},0) + 2F_{c}^{2}]/3$            &               &\\
$R(F)=\sum(||F_{o}|- |F_{c}||)/\sum|F_{o}|
    For F_{o}>4.0\sigma(F_{o})$                      &0.0151        & 0.0167\\
Absolute-Structure parameter Flack's x              &0.50          &0.50\\
$GooF=S=[\sum[w(F_{o}^{2}-F_{c}^{2})^{2}]/(n-p)]^{1/2}$& 1.075        &0.855\\
    $n$\ = number of reflections                       &               &\\
    $p$\ = number of parameters refined                &              &\\
Residual electron density in final                  &-1.0, 0.6(1)  &-0.5,0.8(1)\\
    Difference Fourier map, $e/$\AA$^{3}$                    &              &\\
Max. (shift/s) final cycle                          &0.001        &0.003\\
Average (shift/$\sigma$) final cycle                  &0.000         &0.000\\
\end{tabular}
\end{table}

Crystal data and numerical details on data collection and
refinement are given in Table \ref{roomlowTcomparison}. Final
fractional atomic coordinates, equivalent displacement parameters
and anisotropic displacement parameters are given in Table
\ref{roomt-par-tab} for data at RT and in Table \ref{lowT-par-tab}
for the 90 K data.

\begin{table}[htb]
\caption{Fractional atomic coordinates and Anisotropic Parameters
at RT} \label{roomt-par-tab}
\begin{tabular}{ccccccc}
\hline       Atom &Wyckoff & Schoenflies& x & y & z & $U_{eq}$(\AA) \\
\hline
    Gd & 3 a &D$_{3}$ & 0          & 0          & 0          & 0.00867(5)\\
    Fe & 9 d &C$_{2}$ & 0.21659(5) & 1/3        & 1/3        & 0.00654(9)\\
    O1 & 9 e &C$_{2}$ & 0.1442(2)  & x          & 1/2        & 0.0083(3)\\
    O2 & 9 e &C$_{2}$ & 0.4087(3)  & x          & 1/2        & 0.0132(5)\\
    O3 & 18 f& C$_{1}$ & 0.0254(2)  & 0.2125(2)  & 0.1824(2)  & 0.0095(3)\\
    B1 & 3 b &D$_{3}$ & 0          & 0          & 1/2        & 0.0066(6)\\
    B2 & 9 e &C$_{2}$ & 0.5526(3)  & x          & 1/2        & 0.0077(4)\\
\hline\\
&&&&&&\\
\end{tabular}

\begin{tabular}{ccccccc}
\hline
  Atom & $U_{11}$ & $U_{22}$ & $U_{33}$ & $U_{23}$ & $U_{13}$ & $U_{12}$ \\
\hline
 Gd & 0.00888(8) & 0.00888(8) & 0.00825(9)& 0.0000(-)& 0.0000(-)& 0.00444(4)\\
 Fe & 0.00591(12)&0.00635(14) & 0.00750(18)&0.00009(8)&0.00004(4)&0.00317(7)\\
 O1 & 0.0058(5)  &0.0058(5)   & 0.0109(7)  &0.0014(3) &0.0014(3)&0.0011(6)\\
 O2 & 0.0076(6)  &0.0076(6)   & 0.0182(11) &0.0047(5) &0.0047(5)&0.0009(8)\\
 O3 & 0.0065(5)  &0.0112(5)   & 0.0111(5)  &0.0025(5)       &0.0021(4)&0.0046(4)\\
 B1 & 0.0058(9)  &0.0058(9)   &0.0081(14)  &0.0000(0)       &0.0000(-) &0.0029(4)\\
 B2 & 0.0070(7) & 0.0070(7) &0.0093(8) & 0.0007(4) &       0.0007(4) &0.0036(11)\\
\hline
\end{tabular}
\end{table}

\begin{table}
\caption{Fractional atomic coordinates and Anisotropic Parameters
at T=90~K} \label{lowT-par-tab}
\begin{tabular}{ccccccc}
 \hline       Atom &Wyckoff & Schoenflies& x & y & z & $U_{eq}$(\AA) \\
\hline
 Gd & 3 a &C$_{2}$ & -0.33342(1) & x           & 0           & 0.00406(4)\\
 Fe1& 3 a &C$_{2}$ & 0.11536(5)  & x           & 0           & 0.00360(6)\\
 Fe2& 6 c &C$_{1}$ & -0.21420(6) &-0.54975(4)  & 0.34154(2)  & 0.00366(6)\\
 O1 & 3 b &C$_{2}$ & 0           & -0.07819(15)& 1/6         & 0.0072(3)\\
 O2 & 6 c &C$_{1}$ & -0.5832(2)  & -0.2709(1)  & 0.13774(12) &0.00692(17)\\
 O3 & 6 c &C$_{1}$ & -0.1194(3)  & -0.30445(26)& -0.17980(18)& 0.0057(3)\\
 O4 & 6 c &C$_{1}$ & -0.1467(3)  & -0.36234(16)&0.18479(18)& 0.0058(2)\\
 O5 & 6 c &C$_{1}$ & 0.4755(2)   & 0.1451(2)   & -0.15980(8) & 0.0057(3)\\
 O6 & 3 b &C$_{2}$ & 0.1877(2)   & 0           & 5/6         & 0.0049(3)\\
 O7 & 6 c &C$_{1}$ & -0.5235(3)  & -0.53811(17)&-0.18533(18) & 0.0056(2)\\
 B2a& 6 c &C$_{1}$ & -0.4473(4)  & -0.1201(-)  & 0.15617(14) & 0.0053(3)\\
 B2b& 3 b &C$_{2}$ & 0           & -0.2223(3)  & 1/6         & 0.0049(4)\\
 B1 & 3 b &C$_{2}$ & 0.33204(14) & 0           & 5/6         & 0.0045(5)\\
 \hline\\
 &&&&&&\\
\end{tabular}

\begin{tabular}{ccccccc}
\hline
  Atom & $U_{11}$ & $U_{22}$ & $U_{33}$ & $U_{23}$ & $U_{13}$ & $U_{12}$ \\
  \hline       Gd & 0.00409(6) & 0.00409(7) & 0.00408(7) & 0.00010(2)& 0.00009(1) &0.00210(3)\\
     Fe & 0.00355(10)&0.00355(10) & 0.00355(10)&0.00006(4) &0.00006(4)  &0.00166(10)\\
     Fe & 0.00349(9) &0.00356(11) & 0.00378(9) &0.00013(7) &0.00003(5)  &0.00166(6)\\
     O1 & 0.0081(6)  &0.0054(4)   & 0.0089(4)  &0.00121(19)&0.0024(4)   &0.0040(3)\\
     O2 & 0.0058(3)  &0.0062(3)   & 0.081(3 &0.0012(3)  &0.0014(3)   &0.0025(3)\\
     O3 & 0.0062(4)  &0.0040(5)   & 0.0065(4)  &0.0006(3)  &0.0007(3)   &0.0023(4)\\
     O4 & 0.0054(4)  &0.0045(4)   & 0.0067(4)  &0.0007(3)  &0.0003(3)   &0.0019(4)\\
     O5 & 0.0051(4)  &0.0050(4)   & 0.0063(5) &0.0007(3)  &0.0003(3)   &0.0019(5)\\
     O6 & 0.0044(4)  &0.0058(6)   & 0.0062(5)  &0.0019(5)  &0.0010(2)   &0.0029(3)\\
     O7 & 0.0047(4)  &0.0063(4)   & 0.0064(4)  &0.0008(3)  &0.0001(3)   &0.0033(4)\\
    B2a & 0.0050(6)  &0.0067(7)   &0.0055(5)  &0.0004(5)  &0.0001(4)   &0.0040(4)\\
    B2b & 0.0018(8)  & 0.0061(6)  &0.0055(7)   & 0.0004(3) &0.0008(6)   &0.0009(4)\\
     B1 & 0.0042(8)  & 0.0048(9)  &0.0046(11)  & 0.0009(5) &0.0004(2)   &0.0024(4)\\
\hline
 \end{tabular}
\end{table}

\section{Discussion of the structure}

After a brief introduction describing the main features of
GdFe$_3$(BO$_3$)$_4$, we will focus on the difference between the
RT and the 90~K (LT) structure. First, we will try to understand
the consequences that the structural changes have on the
interpretation of Raman spectra anomalies at the weak first-order
phase transition reported in the literature
 \cite{levi}. Second, we will give a new perspective for the
interpretation of low temperature magnetic data
 \cite{levi,bala}.

The room temperature structure of GdFe$_3$(BO$_3$)$_4$ belongs to
the $R$32 space group (Fig. \ref{structure}a and
\ref{structure}b). Our measurement confirms the structure reported
previously for powder samples
 \cite{Juob} and for several single crystals \cite{Belo,Camp} from
the family RFe$_3$(BO$_3$)$_4$. The structure consists of
alternating layers (parallel to $ab$-plane) of Fe-Gd and BO$_{3}$
groups (see Fig. 1b). The main features of this structure are
already described in literature \cite{Camp}: the BO$_{3}$ groups
are arranged in layers nearly perpendicular to the C$_{3}$ axis
and the Fe atoms are arranged in helicoidal chains parallel to
this axis. Different chains are connected by GdO$_{6}$ and
BO$_{3}$ groups, where each individual BO$_3$ and GdO$_{6}$ group
connects three chains. The distance between Fe atoms in the same
chain ($3.1669(4)$~\AA) is shorter than the Fe-Fe distances for
two nearest chains, that varies along the chain, $4.8308(5)$~\AA,
being the shortest one. The main exchange interaction between
Fe$^{3+}$ is therefore of quasi-1D nature.

\begin{figure}[htb]
\centerline{\includegraphics[width=85mm]{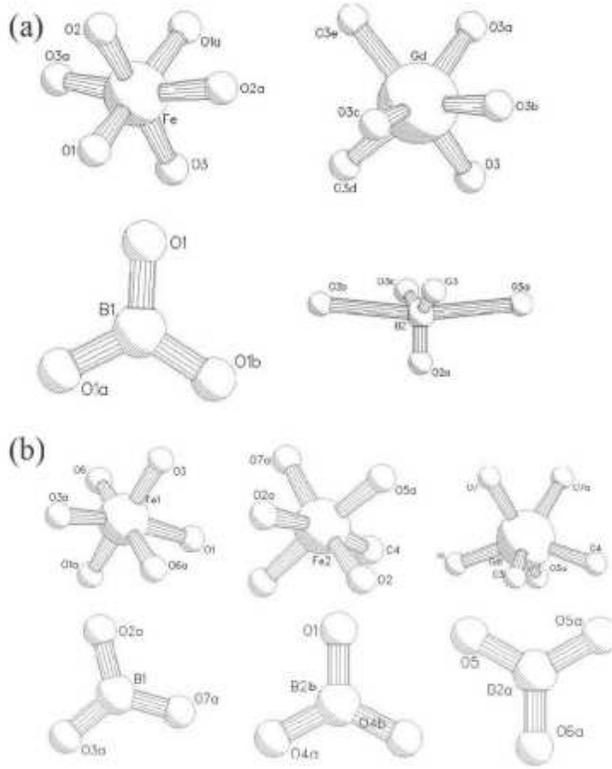}}
\caption{Coordination polyhedra for the RT ($R$32-(a)) and the LT
($P$3$_{1}$21-(b)) structures of GdFe$_{3}$(BO$_{3}$)$_{4}$. At RT
the Fe atoms with C$_{2}$-symmetry are coordinated by three
different types of oxygen atoms. AT 90~K there are two
inequivalent positions for the Fe atoms: Fe1 (the same symmetry as
RT) and Fe2, which is in a general position (surrounded by five
different types of oxygen). At RT the Gd atoms are surrounded by
six oxygens of one type (D$_{3}$-symmetry). At 90~K they are
coordinated by three different types of oxygen atoms
(C$_{2}$-symmetry). For the B atoms there are two kinds of
coordination at RT. B1 (in D$_{3}$-symmetry) is surrounded by
three oxygen atoms of the same type, the B1O$_{3}$-group is thus
an equilateral triangle. The B2 atoms are surrounded by two types
of O, B2O$_{3}$ is an isosceles triangle (C$_{2}$-symmetry). At
90~K there are three kinds of coordinations for the B atoms:
B2bO$_{3}$ and B1O$_{3}$ are isosceles triangles (C$_{2}$
symmetry) and B2aO$_{3}$ is general triangle. } \label{corr}
\end{figure}

\begin{table}[htb]
\caption{Symmetry position of BO$_{3}$ groups, angle between the
group and the \textbf{\textit{c}}-axis and their flatness of them
The flatness is expressed as distance of the B atoms from the
plane defined by the three oxygen ligands.} \label{tilt}
\begin{tabular}{cccccccl}
\multicolumn{4}{c}{Room Temperature } & \multicolumn{4}{c}{Low Temperature } \\
&   Symm. & Angle [degr.]& Flatness &  &   Symm. & Angle [degr.]& Flatness\\
B1 &  D$_{3}$& 90    &   0 & B1  & C$_{2}$ & 87.52(4)& 0.00002(4)\\
B2 &  C$_{2}$ & 84.37(11)  &0 & B2a  & C$_{1}$ & 81.89& 0.0055(11)\\
                  &     &       & & B2b  & C$_{2}$ & 83.55& 0.0000(2)\\
\end{tabular}
\end{table}

Upon lowering the temperature the GdFe$_3$(BO$_3$)$_4$ crystal
reduces the symmetry from  $R$32 to $P$3$_{1}$21, in the trigonal
system. Fig. \ref{corr} shows the coordination polyhedra
(GdO$_{6}$, FeO$_{6}$, and BO$_{3}$) for two different structures
$R$32 and $P$3$_{1}$21. At RT the BO$_{3}$ groups occupy two
inequivalent positions, B1 (D$_{3}$) and B2 (C$_{2}$). At 90~K the
site symmetry of the B1 atoms is reduced to C$_{2}$, whereas the
site symmetry of the B2 atoms differentiates into a B2b (C$_{2}$)
and B2a (C$_{1}$) (see also Table \ref{tilt}). In the LT-phase,
the angle between BO$_{3}$ groups and \textbf{\textit{c}}-axis is
changed. Moreover, in the LT-phase the BO$_{3}$ groups in C$_{1}$
position are considerably distorted and no longer flat (see
Tab.\ref{tilt}).

\begin{figure}
\centerline{\includegraphics[width=85mm]{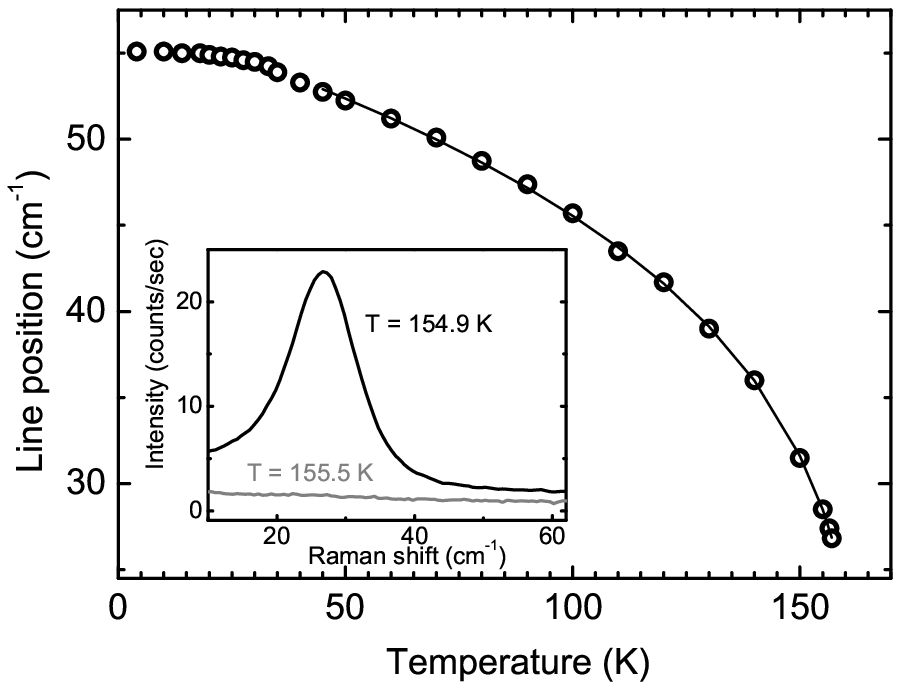}}
\caption{Temperature dependence of Raman frequency of the low
energy mode appearing at the structural phase transition
$T_{\mathrm{s}}$. Insert shows the low-frequency part of Raman
spectrum at two close temperatures before and after phase
transition.} \label{softmode}
\end{figure}

The LT structure is in agreement with the Raman data
\cite{levi,loos}. The group theoretical vibrational analysis,
based on the 90~K structure, shows that, due to the lowering of
the symmetry, new librational modes of the BO$_{3}$ become Raman
active ($R_{x}$ and $R_{y}$). The observed structural change is
compatible with an appearance of the librational $R_{y}$ mode of
the BO$_{3}$ group upon approaching $T_{s}$ from above, and a
subsequent hardening of this mode in the LT phase (See
Fig.\ref{softmode}). Also the anisotropic displacement parameters
(Table \ref{roomt-par-tab}) $U_{33}$ of the oxygen atoms suggest
that the borate groups are relatively free to oscillate around the
$y$-axis.

\begin{figure}
\centerline{\includegraphics[width=50mm]{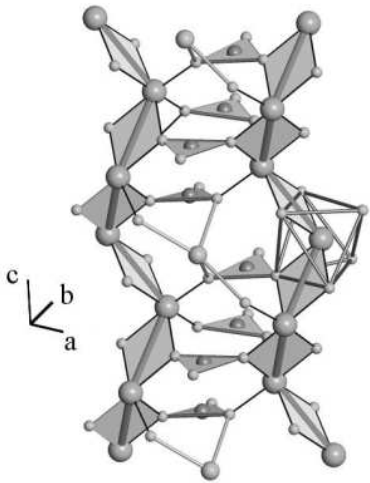}} \caption{Two
Fe-chains and the most important exchange paths: the intra-chain
exchange is via Fe-Fe direct exchange or Fe-O-Fe super-exchange,
while the inter-chain exchange is through Fe-O-Gd-O-Fe and
Fe-O-O-Fe.} \label{exchange}
\end{figure}

Concerning the magnetic structure, the main peculiarity of both
the RT and the 90~K structures is the existence of magnetically
quasi-1D helicoidal iron chains (see Fig. \ref{structure}). The
intra-chain exchange interaction between the Fe ions is expected
to be dominated by Fe-Fe direct exchange and Fe-O-Fe
superexchange, depending respectively on Fe-Fe distance and two
Fe-O-Fe angles (Fig.\ref{exchange}). Nevertheless there are some
significant differences.

At RT all the Fe atoms are in equivalent positions (C$_{2}$). All
Fe-chains are equivalent as are the Fe-O-Fe angles
($102.40^{\circ}$(12) and $103.65^{\circ}$(8)) and Fe-Fe distances
($3.1669(4)$~\AA). Therefore, the exchange interactions between
neighboring iron ions within a chain are also equivalent. At LT
(Fig. \ref{structure}), as shown in Table \ref{tilt}, the BO$_{3}$
groups form one general triangle (gray C$_{1}$ position) and two
isosceles triangles (white and dark, C$_{2}$ position). At RT
"gray" and "white" groups become also equivalent, with C$_{2}$
symmetry, and the "dark" one is a regular triangle with the
D$_{3}$ symmetry. The reduction of the symmetry of the borate
groups changes the surrounding of the Fe atoms (Fig. \ref{corr}),
yielding two inequivalent positions (C$_{2}$ and C$_{1}$).
Therefore, the Fe-Fe distances are different for the two chains:
one is stretched ($3.1828(4)$~\AA) and in the other one is
compressed ($3.1554(4)$~\AA). The angles Fe-O-Fe for the first
chain are $101.24^{\circ}$(5) and $103.71^{\circ}$(9), while those
for the second are $102.46^{\circ}$(6) and $103.91^{\circ}$(8).
Therefore also the intra-chain exchange interaction is different
for the two chains.

The GdO$_{6}$ prism connects three chains, one containing
Fe1-atoms with C$_{2}$ symmetry and two with Fe2-atoms in a
general position. There is only one inequivalent position for the
Gd-ions in both the RT and the 90~K structures. However, the site
symmetry of the Gd-ions changes from D$_{3}$ at RT to C$_{2}$ at
90~K. This confirms the interpretation of the authors of Ref.
 \cite{levi}, who interpreted low-temperature infrared spectra of
Nd$_{0.01}$Gd$_{0.99}$Fe$_{3}$(BO$_{3}$)$_{4}$ in terms of Kramers
doublets of Nd, assuming only one structural position for the Nd
ions.

The detailed structure of the Fe-chains and their interconnection
are shown in Fig. \ref{exchange}. The intra-chain interactions
between Fe-atoms go through two Fe-O-Fe superexchange pathways.
The magnetic inter-chain interaction arises from the Fe-O-O-Fe,
and possibly the Fe-O-Gd-O-Fe superexchange path. The role of the
first superexchange path is important because the substitution of
Gd for non-magnetic Y does not lead to a disappearance of 3D
magnetic ordering. On the contrary, the N\'eel temperature for
YFe$_3$(BO$_3$)$_4$, $T_{\mathrm{N1}}$=40~K, is larger than the
one found for Gd-compound ($T_{\mathrm{N1}}$=37~K), and for
NdFe$_3$(BO$_3$)$_4$ ($T_{\mathrm{N1}}$=30~K) \cite{chuk,loos}.
Considering the different RE$^{3+}$ ionic radii ($0.983$~\AA\ for
Nd, $0.938$~\AA\ for Gd, and $0.900$~\AA\ for Y) it is clear that
the Ne\'el temperature of RFe$_3$(BO$_3$)$_4$ depends strongly on
the ionic radii: a smaller ionic radius results in a higher
$T_{\mathrm{N1}}$. Moreover, $T_{\mathrm{N1}}$ does not seem to be
affected by the spin of the RE. In this sense it is clear that the
main superexchange path of the interaction between different
chains is the Fe-O-O-Fe path, and that a small distortion of this
path changes substantially the magnetic properties of the system.
It is therefore clear that the existence of two nonequivalent Fe
chains with different Fe-Fe distances and Fe-O-Fe angles could
lead to substantially different intra-chain exchange constant for
the two chains \cite{goodenough}, therefore it should be taken
into account for the interpretation of the magnetic properties.

\section{Conclusions}

In summary, we have determined the crystal structure of
GdFe$_3$(BO$_3$)$_4$ at RT and 90~K. At RT GdFe$_3$(BO$_3$)$_4$
exhibits $R$32 structure, in agreement with Refs.
\cite{Juob,Camp}. Below the structural phase transition
($T_{\mathrm{s}}=156$~K) the structure has the $P$3$_{1}$21 space
group. The main difference of LT-structure compared to the RT one
is the lowering of the symmetry and the tilt of the BO$_{3}$
groups. This confirms the interpretation of the Raman spectra of
this phase transition. The main conclusion resulting from the LT
structure determination is the presence of two inequivalent
positions for the Fe atoms giving rise to two different iron
helicoidal chains.

\section{Acknowledgments}
The authors are very grateful to M.N. Popova for valuable
discussions.  This work was partially supported by the Stichting
voor Fundamenteel Onderzoek der Materie (FOM, financially
supported by the Nederlandse Organisatie voor Wetenschappelijk
Onderzoek (NWO)). One of the authors (S.K.) acknowledges the
support of  the Russian Foundation for Basic Research, grant
¹04-02-17346, and the Russian Academy of Sciences under the
Programs for Basic Research.
\bibliography{gfbo}

\end{document}